\documentclass[pra,aps]{revtex4}
\usepackage{graphicx}

\begin{document}

\title{Zigzag spin-S chain near ferromagnet-antiferromagnet transition point}
\author{D.V.Dmitriev}
 \email{dmitriev@deom.chph.ras.ru}
\author{V.Ya.Krivnov}
\affiliation{Joint Institute of Chemical Physics of RAS, Kosygin
str.4, 119991, Moscow, Russia.}
\author{J.Richter}
\affiliation{Institut fuer Theoretische Physik, Universitaet
Magdeburg, P.O. Box 4120, D-39016, Magdeburg, Germany}
\date{}

\begin{abstract}
The properties of the ferromagnetic frustrated spin-S
one-dimensional Heisenberg model in the vicinity of the transition
point from the ferromagnetic to the singlet ground state is
studied using the perturbation theory (PT) in small parameter
characterizing the deviation from the transition point. The
critical exponents defining the behavior of the ground state
energy and spin correlation functions are determined using scaling
estimates of infrared divergencies of the PT. It is shown that the
quantum fluctuations for $s=1/2$ are sufficiently strong to change
the classical critical exponents, while for spin systems with
$s\geq 1$ the critical exponents remain classical. The
dimerization in the singlet phase near the transition point is
discussed.
\end{abstract}

\maketitle

\section{Introduction}

The quantum spin chains with nearest-neighbor (NN) $J_1$ and
next-nearest-neighbor (NNN) $J_2$ interactions have been subject
of numerous studies \cite{review}. The model with both
antiferromagnetic interactions $J_1,J_2>0$ (AF-AF model) is well
studied \cite{Haldane,Tonegawa87,Okamoto,Bursill,Majumdar,White}.
The case of F-AF interactions ($J_1<0,J_2>0$) is less studied.
Though the latter model has been subject of many studies
\cite{Tonegawa89,Chubukov,KO,Aligia,Vekua,Lu}, the complete
picture of the phases of this model as a function of the
frustration parameter $J_2/J_1$ is unclear up to now. An
additional motivation to study this model is related to the fact
that a class of recently synthesized compounds containing $CuO$
chains with edge-sharing $CuO_4$ units are described by the F-AF
zigzag model \cite{Mizuno,Drechsler1,Drechsler2,Hase}. The
$Cu-O-Cu$ angle in these compounds is close to $90^{\circ}$ and
usual antiferromagnetic NN exchange between $Cu$ ions is
suppressed. This means that the sign of $J_1$ can be negative,
while the NNN exchange is antiferromagnetic.

It is well known that there is a critical value $J_2/J_1=-1/4$,
where the transition from the ferromagnetic ground state to the
incommensurate singlet state occurs \cite{Schilling,Hamada}. The
study of the character of this quantum transition is one of the
interesting problem related to the F-AF model. In this paper we
focus on the behavior of the model in the vicinity of the
transition point. We hope that this analysis will be useful for
the study of the properties of the edge-shared copper oxides where
the frustration ratio is close to the critical point. In
particular, for edge-shared cuprate $Li_2ZrCuO_4$ the ratio is
$J_2/J_1\sim -0.28$ and for $Rb_2Cu_2Mo_3O_{12}$ it is
$J_2/J_1\sim -0.37$ \cite{helimagnetism1,helimagnetism2}.

The Hamiltonian of the F-AF model is
\begin{equation}
H=-\sum_{n=1}^{N}(\mathbf{S}_{n}\cdot \mathbf{S}_{n+1}-s^{2})+J
\sum_{n=1}^{N}(\mathbf{S}_{n}\cdot \mathbf{S}_{n+2}-s^{2})
\label{H}
\end{equation}
where we put $J_1=-1$ and $J_2=J$, $s$ is a spin value and
periodic boundary conditions are imposed. The constant shifts in
Eq.(\ref{H}) secure the energy of the ferromagnetic state to be
zero.

Unfortunately, this model is not solved exactly. As was noted
above, the ground state of the model (\ref{H}) is ferromagnetic at
$0<J<1/4$ and it becomes a singlet incommensurate state for
$J>1/4$. Though the transition point between these phases is
$J=1/4$ for any $s$ \cite{Schilling}, the spectra of the model
with $s=1/2$ and $s\geq 1$ in this point are different. For
$s=1/2$ the singlet ground state wave function in the transition
point is known exactly \cite{Hamada,DKO97}. It is degenerate with
the ferromagnetic state\ for any even $N$. For $s\geq 1$ the
singlet ground state wave function is unknown. Finite-size
calculations shows that at $J=1/4$ the singlet state lies slightly
higher than the ferromagnetic level and the energies of the
singlet and the ferromagnetic states are equal in the limit $N\to
\infty $ only.

In the vicinity of the transition point at $0<\gamma \ll 1$
$(\gamma =J-1/4) $ the singlet ground state energy $E_0$ behaves
as $E_0\sim \gamma ^{\beta }$, where $\beta $ is a critical
exponent. The classical approximation gives $\beta =2$. The
spin-wave theory as well as some other approximations
\cite{Bursill,KO} do not change this critical exponent.
Unfortunately, the exact diagonalization of finite chains shows a
complicated irregular size dependence of the ground state energy,
which makes the numerical estimation of the critical exponent
$\beta $ impossible. In the paper \cite{DK06} we conjectured that
for $s=1/2$ strong quantum fluctuations changes the critical
exponent and $\beta =5/3$.

We note that the model (\ref{H}) with $s\geq 1$ has not been
studied before and the critical exponent for these cases is
unknown. In this paper we confirm our conjecture for $s=1/2$ using
scaling estimates of the perturbation theory (PT) in small
parameter $\gamma $. We show also that $\beta =2$ for $s\geq 1$,
though the corresponding factor at $\gamma^2$ is different from
the classical value and it depends on $s$.

One of the most important and open questions in the zigzag model
(\ref{H}) is the possibility of the spontaneous dimerization of
the system in the singlet phase accompanying by a gap in the
spectrum. This problem has been mostly studied in the limit of two
weakly coupled AF $s=1/2$ chains ($J\gg 1 $). The one-loop
renormalization group analysis indicates \cite {Nersesyan,Cabra}
that the gap is open. However, the existence of the gap has not
been verified numerically \cite{Cabra}. On the basis of a field
theory consideration it has been proposed \cite{Itoi} that a
finite gap exists, but it is so tiny that it can not be observed
numerically. On the opposite side of the singlet phase, $J\to
1/4$, there are no any reliable results about the dimerization and
the gap. Strong nonmonotonic finite-size effects do not allow to
study the dimerization numerically.

In order to study the problem of the spontaneous dimerization in
the singlet phase of the model (\ref{H}) close to the transition
point $J=1/4$, we consider the generalization of the model
(\ref{H}) by adding to the Hamiltonian $H$ the perturbation in a
form of dimerization term. Unfortunately, the used special version
of the PT did not give us a rigorous answer about the spontaneous
dimerization in the model (\ref{H}). However, it allowed us to
estimate the critical exponent of the dimer order, in case if the
spontaneous dimerization in the model (\ref{H}) exists. Besides,
it allowed us to obtain the critical exponents of the ground state
energy and the dimer order for the dimerized version of the model
(\ref{H}).

The paper is organized as follows. In Sec.II we present the
scaling estimate of the critical exponent $\beta $ using the PT in
$\gamma $\ for the Hamiltonian (\ref{H}) starting from the singlet
ground state at $J_{c}=1/4$. In Sec.III we analyze the PT for the
Hamiltonian which is transformed to new local axes forming a
spiral structure. We establish the scaling behavior associated
with $\gamma $ and with the pitch angle of the spiral. It is shown
that the critical exponents for the ground state energy are
different for the spins $s=1/2$ and $s\geq 1$. In Sec.IV we study
the problem of the spontaneous dimerization in the model
(\ref{H}). In Sec.V we summarize our results.

\section{Scaling estimate of the critical exponent near the transition point
$J=1/4$}

We are interested in the behavior of the model (\ref{H}) in the
vicinity of the transition point $J_{c}=1/4$. For this aim it is
natural to develop the perturbation theory
\begin{eqnarray}
H &=&H_{0}+V_{\gamma }  \nonumber \\
H_{0} &=&-\sum (\mathbf{S}_{n}\cdot
\mathbf{S}_{n+1}-s^{2})+\frac{1}{4}\sum (
\mathbf{S}_{n}\cdot \mathbf{S}_{n+2}-s^{2})  \nonumber \\
V_{\gamma } &=&\gamma \sum (\mathbf{S}_{n}\cdot \mathbf{S}_{n+2}-s^{2})
\label{H0}
\end{eqnarray}
with a small parameter $\gamma =J-1/4\ll 1$ ($\gamma >0$).

At $\gamma >0$ the ground state of the Hamiltonian $H$ is a
singlet. Since the perturbation $V_{\gamma }$ conserves the total
spin $S^{2}$, the PT to the lowest singlet state $\left| \Psi
_{0}\right\rangle $ of the Hamiltonian $H_{0}$ involves only
singlet excited states. The low-lying singlet excitations at the
transition point have very small energies as shown in Figs.(1) and
(2), where we present finite-size calculations of the energy gap
between the two lowest singlets. These calculations show that the
low-lying singlet excitations have different powers in $N$ for the
cases $s=1/2$ and $s\geq 1$. As it will be shown below this fact
leads to different critical exponents for spin systems with
$s=1/2$ and $s\geq 1$.

The perturbation series for the singlet ground state energy can be
written in a form:
\begin{equation}
E_{0}(\gamma )=\left\langle \Psi _{0}\right| V_{\gamma }+V_{\gamma
}\frac{1}{ E_{0}-H_{0}}V_{\gamma }+\ldots \left| \Psi
_{0}\right\rangle  \label{Eseries}
\end{equation}

Suppose that the low-lying excitations acting in the PT behave as
\begin{equation}
E_{k}-E_{0}\sim N^{-\delta }  \label{dEexp}
\end{equation}

The higher orders of perturbation series contain more dangerous
denominators, and, therefore, have higher powers of the infrared
divergency. Therefore, we use scaling arguments to estimate the
critical exponent for the ground-state energy. Below we will
follow only powers of divergencies and omit all numerical factors.

Suppose that the matrix elements of the perturbation operator
$V_{\gamma }$ between low-lying states involved into the PT at
$N\to \infty $ behave as
\begin{equation}
\left\langle \Psi _{i}\right| V_{\gamma }\left| \Psi _{j}\right\rangle \sim
\gamma N^{1-d}  \label{Vgexp}
\end{equation}
with some exponent $d$.

Looking after powers of infrared divergencies in all orders of the
perturbation series the correction to the ground state energy
takes a form:
\begin{equation}
E_{0}(\gamma )\sim \left\langle \Psi _{0}\right| V_{\gamma }\left| \Psi
_{0}\right\rangle \sum_{m=0}^{\infty }c_{m}x^{m}\sim \gamma N^{1-d}\cdot f(x)
\label{Eseries2}
\end{equation}
where $c_{m}$ are unknown constants and
\begin{equation}
x\sim \frac{\left\langle \Psi _{i}\right| V_{\gamma }\left| \Psi
_{k}\right\rangle }{E_{k}-E_{0}}\sim \gamma N^{\delta +1-d}  \label{x}
\end{equation}
is a scaling parameter, which absorbs the infrared divergencies.

The scaling function $f(x)$ at $x\to 0$ is given by the first
order correction. In the thermodynamic limit ($x\to \infty $) the
behavior of $f(x)$ is generally unknown, but the natural condition
$E_{0}(\gamma )\sim N$ at $ N\to \infty $ requires
\begin{equation}
f(x)\sim x^{\frac{d}{\delta +1-d}}  \label{f(x)}
\end{equation}
and, finally
\begin{equation}
E_{0}(\gamma )\sim -N\gamma ^{\frac{\delta +1}{\delta +1-d}}  \label{Eexp}
\end{equation}

The perturbation series for the lowest excited state $E_{1}(\gamma
)$ has the same form as Eq.(\ref{Eseries}). But a requirement of a
finite mass gap (if any) $m=E_{1}(\gamma )-E_{0}(\gamma )\sim
O(1)$ leads to another critical exponent
\begin{equation}
m\sim \gamma ^{\frac{\delta }{\delta +1-d}}  \label{mexp}
\end{equation}

We note, that for the models in fixed points with a linear
spectrum ($\delta =1$), Eqs.(\ref{x})-(\ref{mexp}) reduce to the
well-known formulae \cite {Cardy}
\begin{eqnarray}
x &=&\gamma N^{2-d}  \nonumber \\
E(\gamma ) &\sim &-N\gamma ^{\frac{2}{2-d}}  \nonumber \\
m &\sim &\gamma ^{\frac{1}{2-d}}  \label{CFT}
\end{eqnarray}
where the exponent $d$ represents the scaling dimension of the
perturbation operator.

However, the transition point $J=1/4$ is not a fixed point. Finite
size calculations for the gap between lowest singlet states give
exponents $\delta =4$ for $s=1/2$ chain and $\delta =3$ for $s\geq
1$ chain (see Figs.(1) and (2)).

\begin{figure*}
\includegraphics{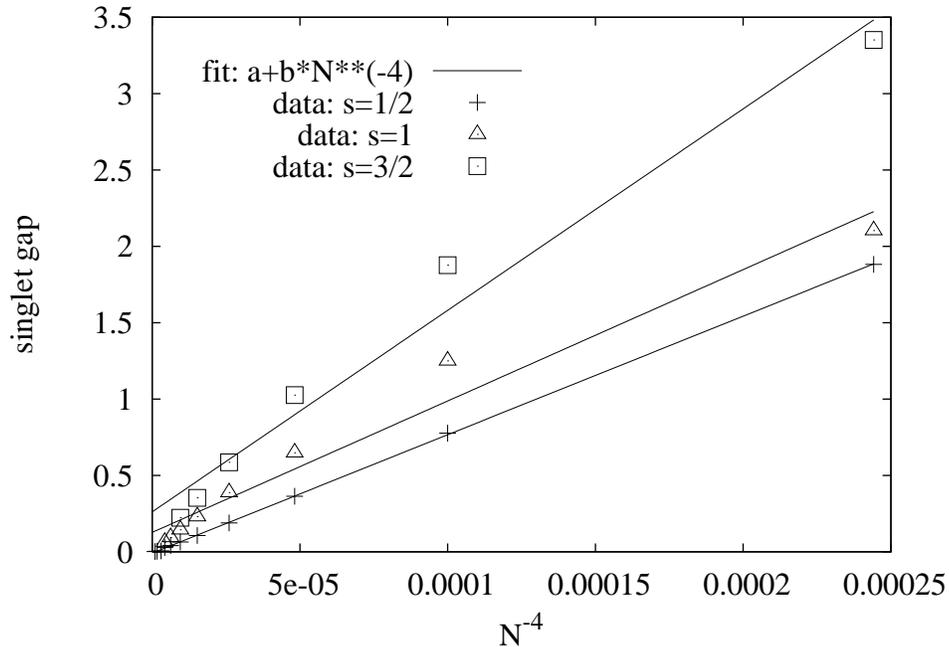}
\caption{The gap between two lowest singlet states of the model
(\ref{H}) at the transition point with different value of spin $s$
vs. $1/N^4$. Only $s=1/2$ curve shows good linear behavior.}
\label{fig_1}
\end{figure*}

\begin{figure*}
\includegraphics{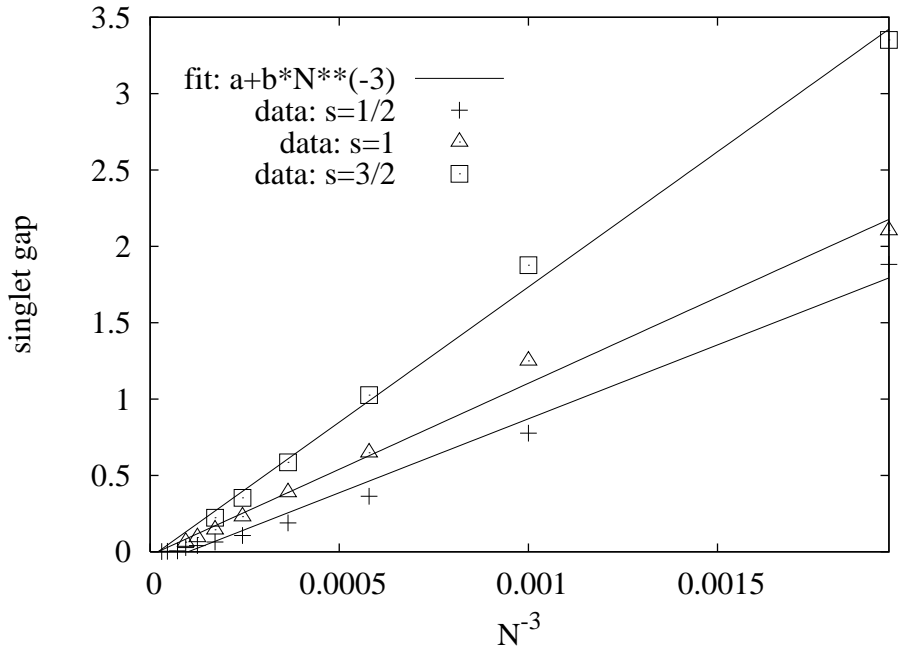}
\caption{The gap between two lowest singlet states of the model
(\ref{H}) at the transition point with different value of spin $s$
vs. $1/N^3$. $s=1$ and $s=3/2$ curves have linear dependence.}
\label{fig_2}
\end{figure*}

In order to determine the value of the exponent $d$, we notice
that the singlet ground state of $H_{0}$ has a spiral ordering at
$N\to \infty $ with a period of the spiral equal to $N$
\begin{equation}
\left\langle \Psi _{0}\right| \mathbf{S}_{n}\cdot \mathbf{S}_{n+l}\left|
\Psi _{0}\right\rangle =s^{2}\cos \frac{2\pi l}{N}  \label{spiral}
\end{equation}

For the case $s=1/2$ this expression is an exact one
\cite{Hamada},\cite{DKO97}, while for $s\geq 1$ we have observed
the spin spiral structure in exact diagonalization of finite-size
systems (the spin correlation function for spin $s=1$ chain of
size $N=20$ is shown in Fig.3). This means that the first order
correction to the ground state energy in $\gamma $ is
\begin{equation}
\left\langle \Psi _{0}\right| V_{\gamma }\left| \Psi _{0}\right\rangle
=-\gamma \frac{\left( 4\pi s\right) ^{2}}{N}  \label{E1singlet}
\end{equation}

\begin{figure*}
\includegraphics{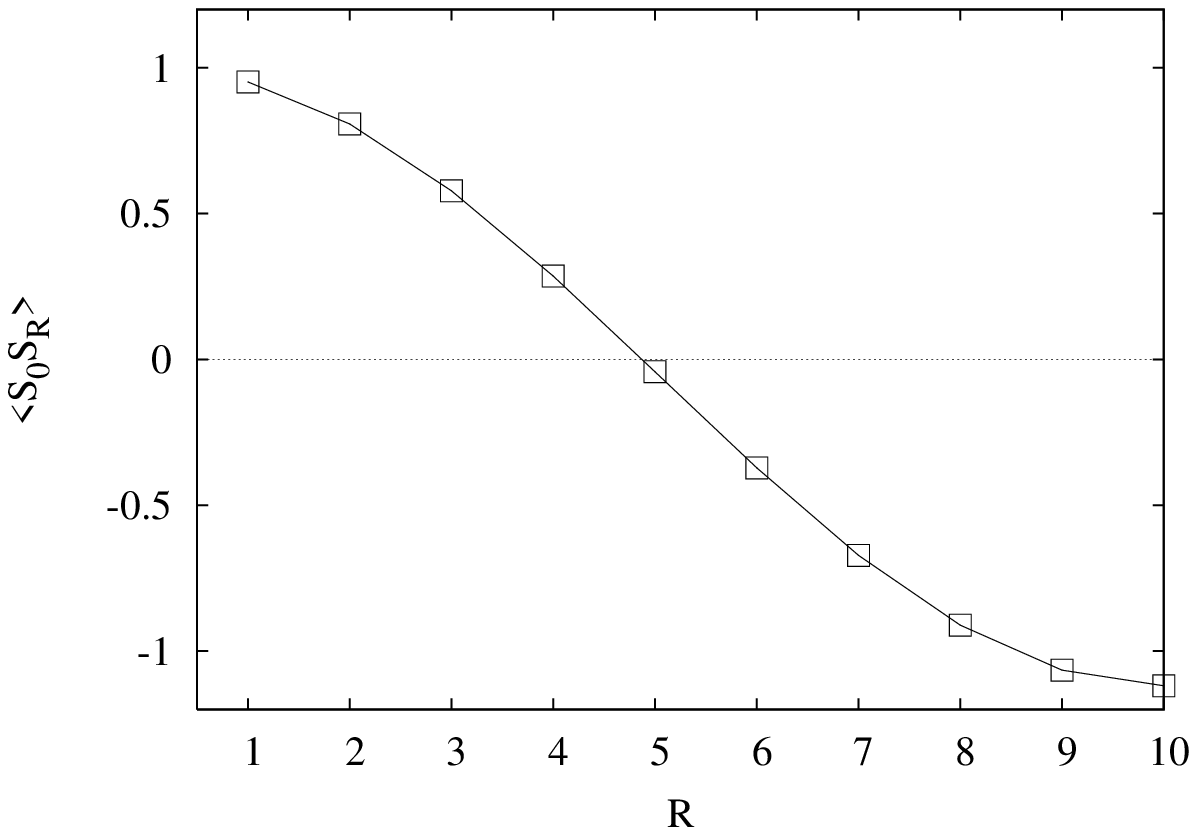}
\caption{The spin correlation function $\left\langle
\mathbf{S}_{n}\cdot \mathbf{S}_{n+l}\right\rangle$ in the lowest
singlet state at the transition point for spin $s=1$ chain of size
$N=20$. The spin spiral structure is obvious.} \label{fig_3}
\end{figure*}

We assume that all matrix elements between any low-lying singlet
states $\left| \Psi _{i}\right\rangle $ and $\left| \Psi
_{j}\right\rangle $ have the same $N$-dependence
\begin{equation}
\left\langle \Psi _{i}\right| V_{\gamma }\left| \Psi _{j}\right\rangle \sim
\frac{\gamma }{N}  \label{Vgsinglet}
\end{equation}
and, therefore, the exponent $d=2$.

Thus, as follows from Eq.(\ref{Eexp}) the critical exponents for
the ground state energy are different for $s=1/2$ and $s\geq 1$
\begin{eqnarray}
E_{0}(\gamma ) &\sim &-N\gamma ^{5/3},\qquad s=1/2  \nonumber \\
E_{0}(\gamma ) &\sim &-N\gamma ^{2},\qquad s\geq 1  \label{Egexp}
\end{eqnarray}
\qquad

For the case $s\geq 1$ the above scaling estimates reproduce the
classical value for the critical exponent of the ground state
energy. But in the special case $s=1/2$ the quantum fluctuations
are strong enough to change the critical exponent. In order to
understand the nature of the difference between $s=1/2$ and $s\geq
1$ systems and determine factors in Eqs.(\ref{Egexp}) in the next
Section we develop a special version of the PT.

\section{Perturbation theory for the transformed Hamiltonian}

Let us start from the classical picture of the ground state of the
model (\ref{H}). In the classical approximation the spins are
vectors which form the spiral structure with a pitch angle
$\varphi $ between neighboring spins. The classical energy per
site
\begin{equation}
E_{\mathrm{cl}}(\varphi )=Ns^{2}\left[ 1-\cos \varphi -J(1-\cos
(2\varphi ) \right]   \label{Ecl}
\end{equation}
is minimized by the angle
\begin{equation}
\varphi _{\mathrm{cl}}=\cos ^{-1}\frac{1}{4J}  \label{phicl}
\end{equation}

The classical ground state energy is
\begin{equation}
E_{\mathrm{cl}}(\varphi _{\mathrm{cl}})=-N\frac{2s^{2}}{J}\gamma ^{2}
\label{Eclphicl}
\end{equation}

Following this picture we transform local axes on $n$-th site by a
rotation about the $Y$ axis by $\varphi n$. This rotation
transforms the original spin wave functions $\left| \psi
_{n}\right\rangle $ to a new basis depending on the angle $\varphi
$
\begin{equation}
\left| \psi _{n,\varphi }\right\rangle =U_{\varphi }\left| \psi
_{n}\right\rangle   \label{rotate}
\end{equation}
where
\begin{equation}
U_{\varphi }=\exp \left( i\varphi \sum_{n=1}^{N}nS_{n}^{y}\right)
\label{Uphi}
\end{equation}
is the rotation operator and $U_{\varphi }^{\dagger
}=U_{-\varphi}$. For finite cyclic systems the pitch angle
$\varphi $ takes quantized values $\varphi _{m}=\frac{2\pi m}{N}$.

Under the unitary transformation $U_{\varphi }$ the Hamiltonian
$H$ takes a form
\begin{eqnarray}
H_{\varphi } &=&U_{\varphi }HU_{-\varphi }=H+V_{\varphi }  \nonumber \\
V_{\varphi } &=&(1-\cos \varphi )\sum \left[
S_{n}^{x}S_{n+1}^{x}+S_{n}^{z}S_{n+1}^{z}\right]   \nonumber \\
&&-J(1-\cos 2\varphi )\sum \left[
S_{n}^{x}S_{n+2}^{x}+S_{n}^{z}S_{n+2}^{z}
\right]   \nonumber \\
&&-\sum [\sin \varphi (S_{n}^{x}S_{n+1}^{z}-S_{n}^{z}S_{n+1}^{x})-J\sin
2\varphi (S_{n}^{x}S_{n+2}^{z}-S_{n}^{z}S_{n+2}^{x})]  \label{Vphi}
\end{eqnarray}

Now let us choose some eigen state $\left| \psi _{n}\right\rangle$
of the Hamiltonian $H$
\begin{equation}
H\left| \psi _{n}\right\rangle =E_{n}\left| \psi _{n}\right\rangle
\label{Heigen}
\end{equation}

The state $\left| \psi _{n}\right\rangle $ is the eigen state of
the Hamiltonian $H$, but not of $H_{\varphi }$. Therefore, if we
develop and exactly calculate the perturbation theory in
$V_{\varphi }$ to this state we arrive to some eigen state $\left|
\psi _{m,\varphi }\right\rangle $ of the Hamiltonian $H_{\varphi
}$
\begin{equation}
H_{\varphi }\left| \psi _{m,\varphi }\right\rangle =E_{m}(\varphi )\left|
\psi _{m,\varphi }\right\rangle   \label{Hphieigen}
\end{equation}
corresponding to, generally speaking, another energy level
$E_{m}(\varphi )\neq E_{n}$. Obviously, the unitary transformation
$U_{\varphi }$ does not change the spectrum. Therefore, the found
energy level $E_{m}(\varphi )$ is also one of the eigen values of
the original Hamiltonian $H$. Thus, taking different values of the
pitch angle $\varphi _{m}=\frac{2\pi m}{N}$ ($m=1\ldots N$) and
developing the PT in $V_{\varphi }$ to some definite eigen state
$\left| \psi _{n}\right\rangle $ of the Hamiltonian $H$ we obtain
a set of $N$, generally different, levels $E_{m}(\varphi )$ of
$H$. So, we do not need to fix the value of $\varphi $ to its
classical value in contrast to the spin-wave approximation.
Instead, we are free to pick out the minimal energy from the set
of the found $N$ levels $E_{m}(\varphi )$. In the thermodynamic
limit, when $\varphi $ becomes continuous variable, this procedure
means the minimization of the found energy $E(\varphi )$ over
$\varphi $.

As a `source' function $\left| \psi _{n}\right\rangle $ of $H$ it
is natural to choose the ferromagnetic state with all spins
pointing up
\begin{equation}
\left| F\right\rangle =\left| \uparrow \uparrow \ldots \uparrow
\right\rangle   \label{Fstate}
\end{equation}

This choice is equivalent to taking the function $\left|
F_{\varphi }\right\rangle =U_{\varphi }\left| F\right\rangle $ as
a probe ground state for the model (\ref{H}). The function $\left|
F_{\varphi }\right\rangle $ has a spiral structure arising in the
classical approximation. The expectation value of the total
$S^{2}$ in this state is \cite{KO}
\begin{equation}
\left\langle F_{\varphi }\right| S^{2}\left| F_{\varphi
}\right\rangle = \frac{N}{2}
\end{equation}

This means that $\left| F_{\varphi }\right\rangle $ is not a pure
singlet state, but contains an admixture of states with $S\neq 0$.
However, it is clear that the weights of states with $S\neq 0$ are
negligible at $N\to \infty $ and we can treat the state $\left|
F_{\varphi }\right\rangle $ as a singlet one.

Since we are interested in the behavior of the model near the
transition point $J=1/4$, it is convenient to represent the
Hamiltonian $H_{\varphi }$ in the form
\begin{equation}
H_{\varphi }=H_{0}+V_{\gamma }+V_{\varphi }  \label{Hphi}
\end{equation}
with $H_{0}$ and $V_{\gamma }$ defined above in Eq.(\ref{H0}) and
to develop the perturbation theory to the ferromagnetic state in
$V=V_{\varphi }+V_{\gamma }$. So, there are two channels
$V_{\varphi }$ and $V_{\gamma }$ in the perturbation theory
characterized by two small parameters $\varphi $ and $\gamma $.
The ferromagnetic state $\left| F\right\rangle $ is the eigen
state of the Hamiltonian $H_{0}$ with the energy $E_{0}=0$ and
also of the perturbation $V_{\gamma }$, but not of $V_{\varphi }$.
The obvious relation $V_{\gamma }\left| F\right\rangle =0$ means
that the perturbation series for the energy contains terms $\sim
\varphi ^{m}\gamma ^{n}$, but does not contain terms $\sim \gamma
^{n}$ without $\varphi $.

At first sight it seems that as a result of the rotation
Eq.(\ref{Uphi}) we obtain more complex Hamiltonian $H_{\varphi }$
and more complicated perturbation theory with two channels. But
the advantage of this method is to construct the perturbation
theory in $V=V_{\varphi }+V_{\gamma }$ to the simple ferromagnetic
state instead of the perturbation theory in $V_{\gamma } $ to very
complicated (and even unknown for $s\geq 1$) lowest singlet state
of $H_{0}$, which was analyzed using scaling arguments and
numerical calculations in the previous section. The fact that we
separate the term $V_{\gamma }$ from $H$ and treat it as the
perturbation does not change our arguments about minimization of
the found expression for energy $E(\varphi ,\gamma )$ over
$\varphi $.

The ground state of the Hamiltonian $H_{0}$ is manifold
degenerate: all the ferromagnetic states $\left|
F_{S_{z}}\right\rangle $ with different total $S_{z}=\sum
S_{n}^{z}$ have zero energy. Therefore, at first we have to split
this degeneration of the ground state with use of secular
equation. It turns out that diagonal elements are proportional to
$\left\langle F_{S_{z}}\right| V\left| F_{S_{z}}\right\rangle \sim
N$, while non-diagonal matrix elements are $\left\langle
F_{S_{z}}\right| V\left| F_{S_{z}^{\prime }}\right\rangle \sim
O(1)$. Therefore, in the thermodynamic limit we can neglect
non-diagonal matrix elements and develop regular perturbation
theory directly to the ferromagnetic state $\left| F\right\rangle
$ with all spins pointing up.

The first-order correction to the energy reproduces the leading
terms of the classical result (\ref{Eclphicl}):
\begin{equation}
E^{(1)}=\left\langle F\right| V_{\varphi }\left| F\right\rangle
=-2Ns^{2}\gamma \varphi ^{2}+Ns^{2}\frac{\varphi ^{4}}{8}  \label{E1}
\end{equation}

The second-order correction to the energy
\begin{equation}
E^{(2)}=\sum_{k}^{\prime }\frac{\left\langle \Psi _{k}\right| V\left|
F\right\rangle ^{2}}{E_{0}-E_{k}}  \label{E2formula}
\end{equation}
relates to a two-magnon states, because operator $V$ (actually
$V_{\varphi }$) have non-zero matrix elements in (\ref{E2formula})
only with the states $\left| \Psi _{k}\right\rangle $ containing
two magnons with total quasi-momentum $Q=0$ and relative
quasi-momentum $k$. Exact calculation of the two-magnon problem
gives for the sum the following result
\begin{equation}
E^{(2)}=-N\frac{3s^{2}\varphi ^{4}}{16(s+1)}  \label{E2}
\end{equation}

This sum converges, because a dangerous denominator
\begin{equation}
\varepsilon _{2}(k)=E_{k}-E_{0}=\frac{s}{2}k^{4}
\end{equation}
for small $k$ is compensated by the matrix elements in a numerator
\begin{equation}
\left\langle \Psi _{k}\right| V_{\varphi }\left| F\right\rangle
=\frac{ 3s^{2}\varphi ^{2}}{4(s+1)}k^{2}  \label{Vphiexp}
\end{equation}

As one can see, the two-magnon spectrum of $H_{0}$ at $Q=0$ and
$k\ll 1$ is simply twice an energy of one magnon $\varepsilon
_{2}(k)=2\varepsilon _{1}(k)$ where
\begin{equation}
\varepsilon _{1}(k)=2s(1-\cos k)-\frac{s}{2}(1-\cos (2k))  \label{Eonemagnon}
\end{equation}
and $\varepsilon _{1}(k)=sk^{4}/4$ at small $k$. So, the low-lying
states of $H_{0}$ with small number of magnons have energies
$\varepsilon _{m}=m\varepsilon _{1}(k)\sim sN^{-4}$, which leads
to infrared divergencies in the next-order corrections to the
energy. Similar to Eq.(\ref{Eseries}) we sum them up using the
scaling arguments.

The PT\ for Eq.(\ref{Hphi}) contains two channels $V_{\gamma }$
and $V_{\varphi }$, which are described by two independent scaling
parameters. In order to determine these scaling parameters one
should estimate large-$N$ behavior of the matrix elements of the
operators $V_{\gamma }$ and $V_{\varphi }$ between low-lying
states $\left| \Psi _{i}\right\rangle $ and $\left| \Psi
_{j}\right\rangle $, acting in the PT. Since the operators
$V_{\gamma }$ and $V_{\varphi }$ create (annihilate) not more than
two magnons, we look after only low-lying states with small number
of magnons and energies
\begin{equation}
\varepsilon _{m}\sim sN^{-4}  \label{Emexp}
\end{equation}

We note that these states are very different from singlet states
(with $N/2$ magnons) presented in Eq.(\ref{Eseries}) and this fact
is crucial.

The diagonal matrix elements for one-magnon states with small quasi-momentum
$k$ behave as
\begin{eqnarray}
\left\langle k\right| V_{\gamma }\left| k\right\rangle &=&-4s\gamma k^{2}
\nonumber \\
\left\langle k\right| V_{\varphi }\left| k\right\rangle
&=&\frac{3}{4} s\varphi ^{2}k^{2}-s\frac{\varphi ^{4}}{4}+4s\gamma
\varphi ^{2} \label{VgVphi1}
\end{eqnarray}
(non-diagonal elements in the one-magnon sector are zero).

For a small number of magnons $m\ll N$ we can treat them as almost
independent, because the interactions between magnons gives only
corrections of the order of magnon density $\rho =m/N$ to the
excitation energies and to the matrix elements. Therefore,
large-$N$ behavior of the matrix elements ($k\sim 1/N$) are
\begin{eqnarray}
\left\langle \Psi _{i}\right| V_{\gamma }\left| \Psi _{j}\right\rangle &\sim
&\gamma sN^{-2}  \nonumber \\
\left\langle \Psi _{i}\right| V_{\varphi }\left| \Psi _{j}\right\rangle
&\sim &\varphi ^{2}sN^{-2}  \label{VgVphi2}
\end{eqnarray}

These formulae are validated by the exact solution of two-magnon
problem.

Now we are ready to identify the scaling parameters of the
perturbations $V_{\gamma }$ and $V_{\varphi }$. Similar to
Eq.(\ref{x}), they are
\begin{eqnarray}
\frac{\left\langle \Psi _{i}\right| V_{\gamma }\left| \Psi _{j}\right\rangle
}{\varepsilon _{m}} &\sim &\gamma N^{2}  \nonumber \\
\frac{\left\langle \Psi _{i}\right| V_{\varphi }\left| \Psi
_{i}\right\rangle }{\varepsilon _{m}} &\sim &\varphi ^{2}N^{2}
\label{VgVphiexp}
\end{eqnarray}

The scaling parameter $\varphi N$ looks natural, because for the
finite cyclic system the pitch angle $\varphi $ is quantized as
$\varphi _{m}=\frac{2\pi m}{N}$.

The infrared divergencies are absorbed by these scaling parameters
so that the divergent part of the perturbation series in both
channels has a form
\begin{equation}
E^{(div)}=\left\langle \Psi _{i}\right| V_{\varphi }\left| \Psi
_{j}\right\rangle \sum_{m,n=0}^{\infty }c_{mn}\left( \varphi N\right)
^{2m}\left( \gamma N^{2}\right) ^{n}  \label{Ediv1}
\end{equation}
with unknown constants $c_{mn}$.

In order to satisfy the thermodynamic relation $E^{(div)}\sim N$,
we rewrite Eq.(\ref{Ediv1}) as
\begin{equation}
E^{(div)}=Ns\varphi ^{5}\sum_{n=0}^{\infty }g_{n}\left( \varphi N\right)
\left( \frac{\gamma }{\varphi ^{2}}\right) ^{n}  \label{Ediv2}
\end{equation}
where
\begin{equation}
g_{n}\left( \varphi N\right) =\sum_{m=0}^{\infty }c_{mn}\left( \varphi
N\right) ^{2n+2m-3}  \label{gn}
\end{equation}
are a set of (generally unknown) scaling functions. They should
converge in the thermodynamic limit $N\to \infty $ to some
constants
\begin{equation}
a_{n}=\lim_{N\to \infty }g_{n}\left( \varphi N\right)  \label{an}
\end{equation}
which are Taylor coefficients of an unknown scaling function
$f\left( \frac{\gamma }{\varphi ^{2}}\right) $:
\begin{equation}
E^{(div)}=Ns\varphi ^{5}\sum_{n=0}^{\infty }a_{n}\left(
\frac{\gamma }{ \varphi ^{2}}\right) ^{n}=Ns\varphi ^{5}f\left(
\frac{\gamma }{\varphi ^{2}} \right)  \label{Ediv3}
\end{equation}

Thus, collecting the converged first-order $E^{(1)}$ and
second-order $E^{(2)}$ corrections with the divergent\ part
$E^{(div)}$ the energy takes the form
\begin{equation}
E=-2Ns^{2}\gamma \varphi ^{2}+Ns^{2}\frac{\varphi
^{4}}{8}\frac{s-1/2}{s+1} +Ns\varphi ^{5}f\left( \frac{\gamma
}{\varphi ^{2}}\right)  \label{E}
\end{equation}

At $\gamma =0$ the estimate of the energy relates to the spectrum
of the Hamiltonian $H_{0}$. One can see that the energy for $s\geq
1$ is $E\sim N\varphi ^{4}$, which for small $\varphi \sim
\frac{1}{N}$ agrees with the numerical estimate $E\sim N^{-3}$\
(see Fig.2). However, in the special case $s=1/2$ the second term
in Eq.(\ref{E}) vanishes and the energy becomes $E\sim N\varphi
^{5}f\left( 0\right) $, which for $\varphi \sim \frac{1}{N}$
agrees again with the numerical estimate $E\sim N^{-4}$ (see
Fig.1). From the positivity of the spectrum of the Hamiltonian
$H_{0}$ we conclude that $f\left( 0\right) >0$.

Now we need to minimize the ground state energy over $\varphi $.
As follows from Eq.(\ref{E}) this procedure is different for
$s=1/2$ and $s\geq 1$. For the case $s=1/2$
\begin{equation}
E=-N\frac{\gamma \varphi ^{2}}{2}+N\frac{\varphi ^{5}}{2}f\left( \frac{%
\gamma }{\varphi ^{2}}\right)  \label{Es12}
\end{equation}

The comparison of powers in $\varphi $ and $\gamma $ of two terms
in Eq.(\ref {Es12}) shows that the minimum of $E$ is reached at
$\varphi _{\min }\sim \gamma ^{1/3}$. Therefore, $f\left(
\frac{\gamma }{\varphi _{\min }^{2}} \right) \to f\left( 0\right)
$ at $\gamma \to 0$ and the expression for $\varphi _{\min }$
takes a form
\begin{equation}
\varphi _{\min }(\gamma )=\left( \frac{2\gamma }{5f(0)}\right) ^{1/3}
\label{phimin12}
\end{equation}

As was shown above $f\left( 0\right) >0$, which justifies
Eq.(\ref{phimin12}). The corresponding minimal energy is
\begin{equation}
E_{\min }=-\frac{0.3}{\left( 2.5f(0)\right) ^{2/3}}N\gamma ^{5/3}
\label{Emin12}
\end{equation}

For the case $s\geq 1$ the minimum is defined by the first two
terms in Eq.(\ref{E}) and
\begin{eqnarray}
\varphi _{\min } &=&\sqrt{8\frac{s+1}{s-1/2}}\sqrt{\gamma }  \nonumber \\
E_{\min } &=&-8Ns^{2}\frac{s+1}{s-1/2}\gamma ^{2}  \label{phimin}
\end{eqnarray}

The last term in Eq.(\ref{E}) gives the correction to the energy
proportional to $\sim Ns\gamma ^{5/2}$.

Thus, we reproduce the critical exponents obtained in Sec.II.
However, this special type of the PT allowed us to determine also
the factor at $\gamma ^{2}$ for the case $s\geq 1$, which at $s\to
\infty $ tends to the classical result Eq.(\ref{Eclphicl}).

According to Eqs.(\ref{phimin12}) and (\ref{phimin}) the pitch
angle $\varphi _{\min }$ has different behavior at $\gamma \to 0$
for $s=1/2 $ and $s\geq 1$. It does not coincide with its
classical value (\ref{phicl}) for any $s$, but it naturally tends
to $\varphi _{\mathrm{cl}}$ at $s\to \infty $. The found non-zero
pitch angle $\varphi _{\min }$ indicates the helical (spiral)
structure of the ground state. Of course, this does not imply the
helical long range order, which is destroyed by strong quantum
fluctuations. Instead, this means an incommensurate behavior of
the spin correlation function and the pitch angle $\varphi _{\min
}$ can be identified with the quasi-momentum $q_{\max }$ at which
the static structure factor takes its maximal value.

\section{Dimerized zigzag model}

In order to study the problem of the spontaneous dimerization in
the zigzag model (\ref{H}) close to transition point $J=1/4$ we
add to the Hamiltonian $H$ the dimerization perturbation
$V_{\alpha }$
\begin{eqnarray}
H_{d} &=&H+V_{\alpha }  \nonumber \\
V_{\alpha } &=&\alpha \sum (-1)^{n}\mathbf{S}_{n}\cdot \mathbf{S}_{n+1}
\label{Hd}
\end{eqnarray}

Then, the behavior of the ground state energy $E_{0}(\alpha
,\gamma )$ of the model (\ref{Hd}) gives us the dimerization order
parameter:
\begin{equation}
p(\alpha ,\gamma )=\frac{1}{N}\left\langle \sum (-1)^{n}\mathbf{S}_{n}\cdot
\mathbf{S}_{n+1}\right\rangle =-\frac{1}{N}\frac{\partial E_{0}(\alpha
,\gamma )}{\partial \alpha }  \label{pdef}
\end{equation}

If $E_{0}(\alpha ,\gamma )\sim -N\alpha p(\gamma )$ at $\alpha \to
0$, then the singlet phase of the model (\ref{H}) is spontaneously
dimerized and $p(\gamma )$ is the corresponding order parameter.

The classical approximation for the model (\ref{Hd}) shows that
the spins form a double-spiral structure defined by two pitch
angles $\varphi $ and $\theta $ so that the rotation angle about
the $Y$ axis on $n$-th site is
\begin{equation}
\varphi _{n}=n\varphi +\frac{(-1)^{n}}{2}\theta   \label{phin}
\end{equation}

The expansion of the classical energy at $(\alpha ,\gamma ,\varphi
,\theta )\ll 1$
\begin{equation}
E_{\mathrm{cl}}(\alpha ,\gamma ,\varphi ,\theta )=Ns^{2}\left( \frac{\varphi
^{4}}{8}+\frac{\theta ^{2}}{2}-2\gamma \varphi ^{2}-\alpha \varphi \theta
\right)  \label{Eclfull}
\end{equation}
is minimized by the angles
\begin{eqnarray}
\varphi _{\mathrm{cl}} &=&\sqrt{8\gamma +2\alpha ^{2}}  \nonumber \\
\theta _{\mathrm{cl}} &=&\alpha \varphi _{\mathrm{cl}}  \label{thetacl}
\end{eqnarray}
which gives the ground state energy at $\alpha ,\gamma \ll 1$
\begin{equation}
E_{\mathrm{cl}}(\alpha ,\gamma )=-\frac{1}{2}Ns^{2}\left( 4\gamma +\alpha
^{2}\right) ^{2}  \label{Eclag}
\end{equation}

As follows from Eq.(\ref{Eclag}) $E_{\mathrm{cl}}(\alpha ,\gamma
)$ vanishes on the line
\begin{equation}
4\gamma +\alpha ^{2}=0  \label{line}
\end{equation}
which determines the transition line between the ferromagnetic and
the singlet phases for the model (\ref{Hd}).

For the case $s=1/2$ the exact singlet ground state on this line
is known \cite{DKO97},\cite{DKOEPJ}. It has double-spiral
long-range order
\[
\left\langle \mathbf{S}_{i}\cdot \mathbf{S}_{i+n}\right\rangle =\frac{1}{4}%
\cos \varphi _{n}
\]
where the angles $\varphi _{n}$ are defined by Eq.(\ref{phin})
with pitch angle $\varphi =\frac{2\pi }{N}$ and small shift angle
between spirals $\theta =\frac{2\pi }{N}\alpha $. It is
interesting that the classical relation $\theta =\alpha \varphi $
(see Eq.(\ref{thetacl})) remains for the strong quantum $s=1/2$
case on the transition line. The dimerization parameter on the
transition line behaves as
\begin{equation}
p_{\mathrm{tr}}=\frac{\varphi \theta }{4}=\frac{\pi ^{2}}{N^{2}}\alpha
\label{pline}
\end{equation}

Though $p_{\mathrm{tr}}\neq 0$ the spontaneous dimerization is
absent on the transition line in the thermodynamic limit.

As follows from Eq.(\ref{pdef}) the classical approximation yields
the dimerization order for the model (\ref{Hd}):
\begin{equation}
p_{\mathrm{cl}}(\alpha ,\gamma )=s^{2}\varphi _{\mathrm{cl}}\theta _{\mathrm{%
cl}}=2s^{2}\alpha \left( 4\gamma +\alpha ^{2}\right)  \label{pcl}
\end{equation}

Eq.(\ref{pcl}) shows that the dimerization vanishes on the
transition line Eq.(\ref{line}), which accords with
Eq.(\ref{pline}), and it vanishes also at $\alpha =0$, which
implies the absence of the spontaneous dimerization for the model
(\ref{H}). Since the classical approximation describes the limit
$s\to \infty $, one can expect that at least in the limit $s\to
\infty $ the spontaneous dimerization in the model (\ref{H}) is
absent.

Following the classical picture we transform the local axes on
$n$-th site by a rotation about the $Y$ axis by angle $\varphi
_{n}$ as written in Eq.(\ref {phin}), but not fixing $\varphi $
and $\theta $ to their classical values. Under this unitary
transformation the Hamiltonian $H_{d}$ (\ref{Hd}) takes a form
\begin{equation}
H_{\varphi ,\theta }=H_{0}+V(\alpha ,\gamma ,\varphi ,\theta )
\label{Hdimer}
\end{equation}
where the perturbation $V(\alpha ,\gamma ,\varphi ,\theta )$ has a
very cumbersome form and we do not present it here.

Similar to the analysis done in Sec.III we develop PT in $V(\alpha
,\gamma ,\varphi ,\theta )$ to the fully polarized state
(\ref{Fstate}). The first order in $V$ exactly reproduces the
classical energy (\ref{Eclfull})
\begin{equation}
E^{(1)}=E_{\mathrm{cl}}(\alpha ,\gamma ,\varphi ,\theta )  \label{dimerE1}
\end{equation}

The second order correction to the ground state energy gives
\begin{equation}
E^{(2)}=-Ns^{2}\left( \frac{3\varphi ^{4}}{16(s+1)}+\frac{(\theta -\alpha
\varphi )^{2}}{2}\right)  \label{dimerE2}
\end{equation}

As one can see the terms containing the angle $\theta $ in the
first order $E^{(1)}$ are exactly compensated by the contributions
of the second order $E^{(2)}$. This result is rather unexpected.
The classical approximation corresponds to the limit $s\to \infty
$ and it would seem that the quantum effects will give relative
corrections $\sim s^{-1}$ to the energy. However, in this case the
quantum corrections have the same order in $s$ as the classical
energy.

The next-order corrections contain infrared divergencies and we
treat them using the scaling arguments similar to that done in
Sec.III. The analysis shows that the most divergent parts of the
PT are accumulated in the following scaling parameters:
\begin{eqnarray}
x_{\alpha } &\sim &\alpha N  \nonumber \\
x_{\varphi } &\sim &\varphi N  \nonumber \\
x_{\gamma } &\sim &\gamma N^{2}  \label{scaling}
\end{eqnarray}

It turns out that the angle $\theta $ is not accompanied by the
infrared divergencies and, therefore, it does not form a scaling
parameter. After the algebraic manipulations with the divergent
series of the PT similar to Eq.(\ref{Ediv1})-(\ref{Ediv3}), the
main contribution of the next-order corrections to the ground
state energy at $N\to \infty $ takes a form
\begin{equation}
E^{(div)}=Ns\varphi ^{5}f\left( \frac{\gamma }{\varphi
^{2}},\frac{\alpha }{\sqrt{\gamma }}\right)  \label{dimerEdiv}
\end{equation}
where $f\left( \frac{\gamma }{\varphi ^{2}},\frac{\alpha
}{\sqrt{\gamma }} \right) $ is unknown scaling function of two
scaling variables. The angle $\theta $ does not contribute to the
most divergent parts of the PT Eq.(\ref {dimerEdiv}), because it
is not accompanied by the infrared divergencies.

Collecting the corrections $E^{(1)}$ and $E^{(2)}$ with the
scaling\ part $E^{(div)}$ we obtain the leading terms for the
ground state energy:
\begin{equation}
E=-\frac{1}{2}Ns^{2}(4\gamma +\alpha ^{2})\varphi ^{2}+Ns^{2}\frac{\varphi
^{4}}{8}\frac{s-1/2}{s+1}+Ns\varphi ^{5}f_{s}\left( \frac{\gamma }{\varphi
^{2}},\frac{\alpha }{\sqrt{\gamma }}\right)  \label{dimerE}
\end{equation}

As follows from Eq.(\ref{dimerE}) the leading terms do not contain
the angle $\theta $. In fact, we have checked that the energy does
not contain terms up to $\sim \theta ^{4}$. This result is not
surprising. In general, the PT in $V(\alpha ,\gamma ,\varphi
,\theta )$ for the ferromagnetic state results in the energy
$E(\alpha ,\gamma ,\varphi ,\theta )$ depending on $\theta $. On
the other hand, the spectra of the Hamiltonians $H_{d}$ (\ref{Hd})
and $H_{\varphi ,\theta }$ (\ref{Hdimer}) coincide and the
eigenvalues $E_{n}(\alpha ,\gamma )$ of both Hamiltonians do not
depend on $\theta $ and $\varphi $. Therefore, for any values of
$\theta $ and $\varphi $ the PT leads to one of the determinate
levels $E_{n}(\alpha ,\gamma )$ of the Hamiltonian $H$. The pitch
angle $\varphi $ is quantized as $\varphi _{n}= \frac{2\pi n}{N}$
for finite $N$, and the PT with different $\varphi _{n}$ leads to
generally different levels $E_{n}(\alpha ,\gamma )$. At the same
time, in contrast to the pitch angle $\varphi $ the angle $\theta
$ is a continuous variable even for finite $N$. Therefore, the
continuity condition of\ the dependence $E(\alpha ,\gamma ,\varphi
,\theta )$ on $\theta $ implies that the PT in $V(\alpha ,\gamma
,\varphi ,\theta )$ with any value of $\theta $ leads to the same
energy level as at $\theta =0$. In other words, the obtained in
the PT ground state energy does not depend on $\theta $.

This fact is an argument for the absence of the spontaneous
dimerization in the zigzag model (\ref{H}). Really, the PT with
any value of $\theta $ brings to the same state as it does at
$\theta =0$. But at $\theta =0$ and $\alpha =0$ the PT in
$V(\alpha ,\gamma ,\varphi ,\theta )$ reduces to the PT
(\ref{Vphi}) in $V_{\varphi }$ considered in Sec.III. There are no
terms in the perturbation $V_{\varphi }$ which break translational
symmetry and can potentially lead to the dimer order.

However, the above arguments do not prove the absence of the
spontaneous dimerization in the model (\ref{H}). The rigorous
method is to calculate the dimer order parameter directly from
Eq.(\ref{pdef}), which we follow below.

The minimization of the ground state energy over $\varphi $ is
performed in the same manner as was done in Sec.III. For the case
$s=1/2$ the second term in Eq.(\ref{dimerE}) disappears. The
comparison of powers in $\varphi $ of two terms in
Eq.(\ref{dimerE}) shows that the minimum of $E$ is reached at
$\varphi _{\min }\sim (4\gamma +\alpha ^{2})^{1/3}$. Therefore, we
substitute $f\left( \frac{\gamma }{\varphi _{\min
}^{2}},\frac{\alpha }{\sqrt{\gamma }} \right) \to f\left(
0,\frac{\alpha }{\sqrt{\gamma }}\right) $ at $\gamma \to 0$ and
the expression for $\varphi _{\min }$ becomes
\begin{equation}
\varphi _{\min }(\alpha ,\gamma )=\left( 4\gamma +\alpha ^{2}\right)
^{1/3}g(\eta )  \label{dimerphi12}
\end{equation}
where $\eta =\frac{\alpha }{\sqrt{\gamma }}$ and $g(\eta )=\left[ 10f\left(
0,\eta \right) \right] ^{-1/3}$.

The corresponding minimal energy is
\begin{equation}
E_{\min }=-\frac{3N}{40}\left( 4\gamma +\alpha ^{2}\right) ^{5/3}g^{2}(\eta )
\label{dimerEmin12}
\end{equation}

For the case $s\geq 1$ the energy minimum is defined by the first
two terms in Eq.(\ref{dimerE}) and
\begin{eqnarray}
\varphi _{\min } &=&\sqrt{2\frac{s+1}{s-1/2}}\sqrt{4\gamma +\alpha ^{2}}
\nonumber \\
E_{\min } &=&-Ns^{2}\frac{s+1}{2s-1}\left( 4\gamma +\alpha ^{2}\right)
^{2}+Ns\left( 4\gamma +\alpha ^{2}\right) ^{5/2}g_{s}\left( \eta \right)
\label{dimerEmin}
\end{eqnarray}
where $g_{s}\left( \eta \right) =\left[ \frac{2\left( s+1\right)
}{s-1/2} \right] ^{5/2}$ $f_{s}\left( \frac{\gamma }{\varphi
_{\min }^{2}},\eta \right) $. So, the leading term in the ground
state energy for $s\geq 1$ is determined by the regular parts of
the PT, while the scaling part (the last term in
Eq.(\ref{dimerEmin})) gives only small correction to the energy.

We see that the difference in the critical exponents for the cases
$s=1/2$ and $s\geq 1$ remains for more general dimerized model
(\ref{Hdimer}) as well. The pitch angle $\varphi _{\min }$ and the
ground state energy $E_{\min }$ naturally vanish on the transition
line Eq.(\ref{line}) for both cases $s=1/2$ and $s\geq 1$.

The dimerization of the model (\ref{Hdimer}) is defined as a
derivative of the energy with respect to $\alpha $ (\ref{pdef}).
As follows from Eqs.(\ref {dimerEmin12}),(\ref{dimerEmin}) the
dimerization at $\gamma =0$ appears with critical exponents
\begin{eqnarray}
\left. p\right| _{\gamma =0} &\sim &\alpha ^{7/3},\qquad s=1/2  \nonumber \\
\left. p\right| _{\gamma =0} &\sim &\alpha ^{3},\qquad s\geq 1  \label{pg0}
\end{eqnarray}

As for the model (\ref{H}) ($\alpha =0$), the dimerization depends
on the behavior of the scaling functions $f_{s}\left( \eta \right)
$ at small $\eta $. There are two possible scenarios. First, the
expansion of $f_{s}\left( \eta \right) $ at $\eta \to 0$ is
$f_{s}\left( \eta \right) =a+b\eta ^{\mu }$ with some constants
$a$ and $b$ and $\mu >1$, so that $f_{s}^{\prime }\left( 0\right)
=0$. In this case the dimer parameter is zero for the model
(\ref{H}). Second, $f_{s}\left( \eta \right) =a+b\eta $ at $\eta
\to 0$ and $f_{s}^{\prime }\left( 0\right) =b$. For this case the
translation symmetry of the zigzag model (\ref{H}) is
spontaneously broken and dimer long-range order $p(\gamma )$
appears as
\begin{eqnarray}
\left. p\right| _{\alpha =0} &\sim &\gamma ^{7/6},\qquad s=1/2  \nonumber \\
\left. p\right| _{\alpha =0} &\sim &\gamma ^{2},\qquad s\geq 1  \label{pa0}
\end{eqnarray}

Here the critical exponent for the dimerization (\ref{pa0}) for
$s\geq 1$ comes not from the leading term in Eq.(\ref{dimerEmin}),
but from the scaling correction (the last term).

Unfortunately, we do not have any information about the behavior
of the scaling functions $f_{s}\left( \eta \right) $. Therefore,
we can only state that if the zigzag model (\ref{H}) is in the
dimerized singlet phase at $\gamma >0$, then the critical
exponents for the dimer LRO are given by Eqs.(\ref{pa0}).

\section{Summary}

We have studied the frustrated Heisenberg chain with the nearest
ferromagnetic and the next-nearest neighbor antiferromagnetic
exchange interactions. It was shown that the behavior of the model
in the vicinity of the transition point between the ferromagnetic
and the singlet phases depends on the value of the spin. For
$s=1/2$ the critical exponent characterizing the behavior of the
energy is $\beta =5/3$ in contrast to the `classical' exponent
$\beta =2$ for $s\geq 1$. This difference is a result of different
finite-size dependencies of the spectrum at the transition point
$\gamma =0$ for the cases $s=1/2$ and $s\geq 1$. The pitch angles
characterizing the incommensurate behavior of the spin correlation
functions are different for $s=1/2$ and $s\geq 1$, too. In
particular, the pitch angle $ \varphi $ of the spiral is
proportional to $\gamma ^{1/3}$ for $s=1/2$ and to $\gamma ^{1/2}$
for $s\geq 1$. It means that the considered model with $s=1/2$ is
special and the quantum effects for this value of $s$ are the most
strong.

One more intriguing question is related to the existence of the
spontaneous dimerization in the singlet phase. In order to study
this problem, we added to the Hamiltonian $H$ the dimerization
term and treated it as a perturbation. Unfortunately, the used
special version of the PT did not give us a rigorous answer about
the spontaneous dimerization in the singlet phase. Instead, under
assumption of the existence of the spontaneous dimerization, the
PT allowed us to estimate the critical exponent of the dimer order
parameter. Besides, using the special version of the PT we obtain
the critical exponents of the ground state energy and the dimer
order for the dimerized version of the model (\ref{H}).

\begin{acknowledgments}
We would like to thank S.-L. Drechsler and R. O. Kuzian for
helpful discussions. D.D. was supported by INTAS YS Grant Nr. 05–
109–4916 and President of RF Grant MK-3987.2005.2. The numerical
exact diagonalization for finite systems  were performed using the
J.Schulenburg's {\it spinpack}.
\end{acknowledgments}

\end{document}